\documentclass[prx,twocolumn,floatix,amssymb,amsmath, superscriptaddress,longbibliography]{revtex4-2}
\usepackage{graphicx}
\usepackage{epsfig}
\usepackage{amsmath,amssymb,amsthm}
\usepackage[utf8]{inputenc}
\usepackage[dvipsnames]{xcolor}
\usepackage[colorlinks, urlcolor=blue, linkcolor=red]{hyperref}
\usepackage{soul,xcolor}

\setstcolor{red}
\newcommand{\beqa}{\begin{eqnarray}}
\newcommand{\eeqa}{\end{eqnarray}}

\renewcommand{\boxed}[2]{\textcolor{#1}{%
\tikz[baseline={([yshift=-1ex]current bounding box.center)}] \node [rectangle, minimum width=1ex,rounded corners,draw] {\normalcolor\m@th$\displaystyle#2$};}}

\newcounter{appsection}
\newcounter{appsubsection}[appsection]

\usepackage{physics}
\usepackage{bookmark}
\usepackage[normalem]{ulem}

\newcommand\redsout{\bgroup\markoverwith{\textcolor{red}{\rule[0.5ex]{2pt}{2pt}}}\ULon}
\newcommand\bluesout{\bgroup\markoverwith{\textcolor{blue}{\rule[0.5ex]{2pt}{2pt}}}\ULon}
\newcommand\greensout{\bgroup\markoverwith{\textcolor{green}{\rule[0.5ex]{2pt}{2pt}}}\ULon}

\newtheorem*{result*}{Main Result}
\newtheorem*{theo*}{Theorem}
\newtheorem{theo}{Theorem}

\begin{document}

\title{Privacy in Distributed Quantum Sensing with Gaussian Quantum Networks}

\author{Uesli Alushi}
\email{uesli.alushi@aalto.fi}
\affiliation{Department of Information and Communications Engineering, Aalto University, Espoo 02150, Finland}
\author{Roberto Di Candia}
\email{rob.dicandia@gmail.com}
\affiliation{Department of Information and Communications Engineering, Aalto University, Espoo 02150, Finland}
\begin{abstract}

We study the privacy properties of distributed quantum sensing protocols in a Gaussian quantum network, where each node encodes a parameter via a local phase shift. We first show that perfect privacy and optimal precision are jointly achievable using specifically tailored multimode photon-number correlated states. We then consider Gaussian states, which are experimentally less demanding as they can be implemented using only linear optics and two-photon parametric processes. Focusing on fully symmetric Gaussian states, we show that for networks with more than two nodes, perfect privacy can be achieved only asymptotically, in the limit of large photon numbers. However, we show that optimized fully-symmetric Gaussian states enable improved privacy levels while maintaining near-optimal sensing performance.  We also show that local homodyne detection is essentially optimal, achieving quadratic scaling of precision with the total number of photons. We further analyze the impact of thermal noise in the preparation stage on both privacy and estimation precision. Our results pave the way for the development of practical, private distributed quantum sensing protocols in continuous-variable quantum networks.
\end{abstract}

\maketitle
\section{Introduction}
\begin{figure}[t!]
    \centering
\includegraphics[width=.48\textwidth]{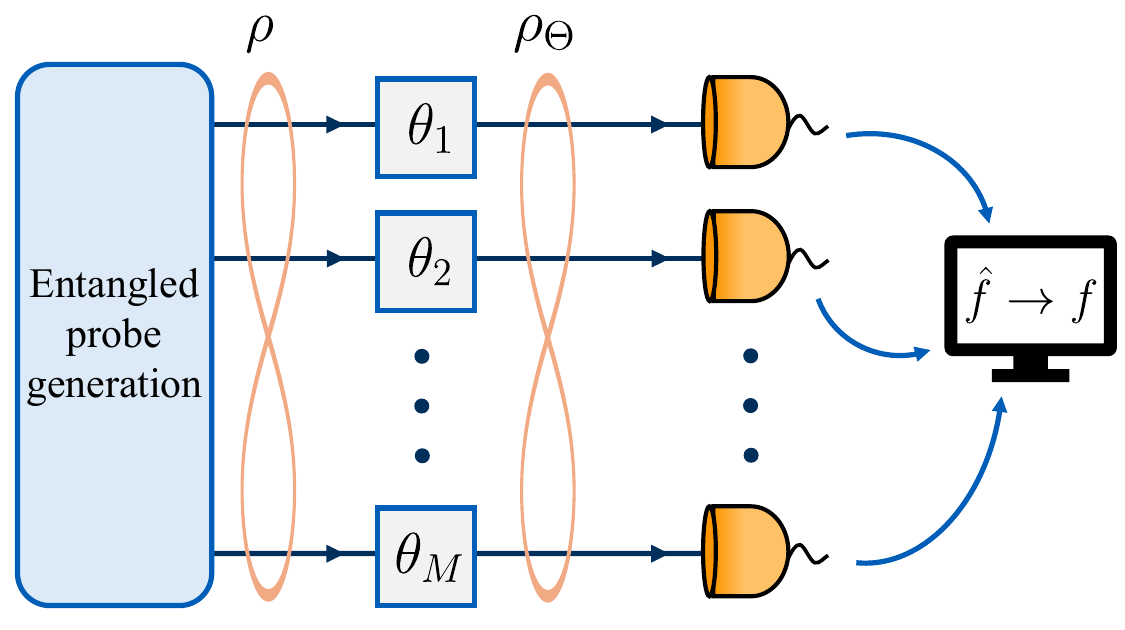}
    \caption{{\bf Scheme of a DQS protocol.} An entangled probe state $\rho$ is prepared and distributed among $M$ users. Each user encodes an unknown parameter $\theta_i$ onto their respective share, resulting in the encoded state $\rho_{\Theta}$. The latter is then locally measured. Finally, a linear combination of the unknown parameters, $f$, is inferred via an unbiased estimator $\hat f$.}
    \label{Fig1}
\end{figure}

Quantum metrology exploits quantum resources to enhance the precision of parameter estimation tasks beyond what is achievable with classical strategies~\cite{Giovannetti2004,Giovannetti2011, PRXQuantum.6.020351, PRXQuantum.6.020301, shi2025quantum, PhysRevLett.133.040801}. These advantages apply not only to single-parameter estimation but also extend to multi-parameter scenarios~\cite{Multiparameter2016,Adesso2018,Liu2020}. In recent years, increasing interest in quantum networks~\cite{chiribella2009,Liu2025} has led to the emergence of distributed quantum sensing (DQS) as a promising framework, wherein multiple spatially separated sensors share entangled probe states to jointly estimate global parameters, such as the mean, with enhanced precision~\cite{Proctor2018,Shapiro2018,Alexey2018,Zhang2021,Malia2022,kim2024distributed,Tag2025}. DQS has found applications in tasks such as global clock synchronization~\cite{Komar2014,Komar2016,Ullah2020,Dai2020,Nande2023,Tang2023} and phase imaging~\cite{Ian2013,Guo2020,Xia2020}. Let us consider a simple DQS scheme, as illustrated in Fig.~\ref{Fig1}. In this quantum network, each user encodes an unknown parameter onto their share of a globally distributed quantum state. The encoded state is then measured, and the outcomes are classically processed to estimate a global parameter expressed as a linear combination of the local phases.

A potential challenge for quantum networks is the risk of information leakage, where untrusted nodes may cooperate to extract information about the locally encoded parameters. To address this concern, the notion of privacy has recently been introduced in the context of quantum networks~\cite{Markham2019} and further developed for distributed quantum sensing protocols~\cite{Markham2022,Markham2025,Paris2025}, with the first experimental demonstrations carried out using discrete-variable systems~\cite{Ho2024}. Here, privacy is understood as the condition in which each party has access only to information about the global target function and their own locally encoded parameter, while being unable to infer information about the parameters of other parties. While the definition of privacy in this setting is general and applicable across different physical platforms, detailed analyses so far have primarily focused on discrete-variable systems. 

In this work, we analyze the privacy properties of a distributed quantum sensing protocol that employs continuous-variable (CV) states as a probe to estimate a linear function of local phase-shifts. We first find that specifically tailored multimode photon-number-correlated states ensure perfect privacy and maximal precision. We then focus on Gaussian states, which are generally less experimentally demanding to generate, as they require at most two-photon parametric processes. For this aim, we focus on fully-symmetric Gaussian (FSG) states~\cite{Adesso2004,Serafini2005,AlessioSerafini}, which are known to be optimal among Gaussian states for estimating the mean of locally encoded phases~\cite{Shapiro2018,Oh2020}. Notably, for these states, collective measurements at the output are not required if the encoded parameter is a displacement or a phase shift; local homodyne detection at each site is essentially optimal to estimate the global parameter efficiently~\cite{Shapiro2018,Zhang2021}. 
We demonstrate that FSG states achieve perfect privacy only asymptotically, in the limit of large photon numbers, and that optimal privacy and precision cannot be attained simultaneously. However, we find FSG states that maximize privacy, incurring a small trade-off in precision, thereby mitigating the need for large entanglement across the network.

\section{Distributed quantum sensing}
\subsection{The protocol}
Consider a parameter \(\theta\) encoded into a quantum state \(\rho\) via a quantum operation \(\Lambda_\theta\), such that \(\rho_\theta=\Lambda_\theta\rho\Lambda^\dagger_\theta\). A standard quantum sensing protocol aims to estimate the parameter \(\theta\) using a specific input probe state \(\rho\) and a suitable measurement strategy. Given an unbiased estimator \(\hat{\theta}\), the estimation error is quantified by its variance \(\Delta^2\hat{\theta}\), which is lower-bounded by the quantum Cramér–Rao bound \(\Delta^2\hat{\theta}\geq1/nF\)~\cite{Paris2009}. Here, \(F\) denotes the quantum Fisher information, and \(n\) is the number of independent repetitions of the protocol. This bound can be saturated for sufficiently large $n$, e.g., by the maximum-likelihood estimator.

Distributed quantum sensing operates in a different, more general setting, see Fig.~\ref{Fig1}. In this context, an \(M-\)mode input probe is shared by a server across a quantum network among \(M\) different users. Each of these users encodes a single parameter \(\theta_i\) onto their portion of the shared state via a local quantum operation \(\Lambda_{\theta_i}\). For instance, in this paper, we consider phase shifts, each generated by the local Hamiltonian ${G_i=a^\dag_i a_i}$. As a result, we now have a vector of unknown parameters \(\Theta=(\theta_1,\theta_2,...,\theta_M)^T\). The task is to estimate a linear combination of these parameters \(f=w^T\Theta=\sum_{i=1}^{M}w_i\theta_i\), where \(w=(w_1,w_2,...,w_M)^T\) is the weight vector. We assume that all weights are strictly positive and, without loss of generality (w.l.o.g.), that they are normalized as \(\|w\|_1=\sum_{i=1}^Mw_i=1\). The estimation error for a locally unbiased estimator $\hat f$ is bounded by~\cite{Paris2009,Zhang2021}: 
\begin{equation}\label{CRbound}
    \Delta^2\hat{f}\geq \frac{w^T{\bf F}^{-1}w}{n}\equiv \frac{1}{n\xi}\,,
\end{equation}
where $\mathbf{F}$ is the quantum Fisher information matrix (QFIM), $n$ is the number of independent repetitions of the protocol. Here, the inverse operation should be understood as a pseudoinverse when the QFIM has zero eigenvalues. However, if ${{\bf F}}$ is singular, then  Eq.~\eqref{CRbound} is saturable only if $w$ is in the range of ${\bf F}$. The estimator $\hat f$ is local, in the sense that it requires prior information about the parameters, namely an a priori mean value and a small enough variance. In what follows, we assume w.l.o.g. that the a priori mean is zero. We refer to ${\xi\equiv (w^T{\bf F}^{-1}w)^{-1}}$ as the estimation precision, i.e., the inverse of the estimation variance for a single run of the protocol. 

The entries of the QFIM are~\cite{Adesso2018}
\begin{equation}
    {\rm F}_{jk}=\frac{1}{2}{\rm Tr}(\rho_\Theta\{L_{j},L_k\})\,,
\end{equation}
where \(L_j\) is the symmetric logarithmic derivative associated to the parameter \(\theta_j\), defined implicitly by \(\partial_{\theta_j}\rho_\Theta=(L_j\rho_\Theta+\rho_\Theta L_j)/2\).

\subsection{Privacy in distributed quantum sensing}
A distributed sensing protocol is said to exhibit perfect privacy when each user has access solely to information about their own local parameter and the global estimate of the target function of parameters~\cite{Markham2022,Markham2025,Paris2025}. In other words, while all users collaborate in the estimation process, none gains any information about the individual parameters held by the others. The privacy level of a distributed sensing protocol can be quantified by the privacy parameter~\cite{Markham2025,Paris2025}
\begin{equation}\label{privacydefinition}
    \mathcal{P}=\frac{1}{\|w\|_2^2}\frac{w^T{\bf F}w}{{\rm Tr(\bf F)}}\,,
\end{equation}
which depends on the particular probe state used. Perfect privacy is achieved only if \({{\bf F}\propto ww^T}\). Indeed, the privacy is maximal when ${\bf F}$ is rank one with eigenvector $w$. Notice that the eigenvectors of the QFIM represent linear combinations of parameters to which the sensing protocol is sensitive, and the corresponding eigenvalues determine the estimation precision for each combination~\cite{Liu2020,Markham2025}. If the QFIM has only one non-zero eigenvalue, then only the linear combination aligned with its corresponding eigenvector can be estimated with finite precision, while all orthogonal combinations become impossible to estimate, thereby ensuring perfect privacy (\(\mathcal{P}=1\)). 

 We emphasize that the notion of privacy considered here is defined in terms of the QFIM, which has an operational meaning in terms of locally unbiased estimators, via the Cramér-Rao bound in Eq.~\eqref{CRbound}. It assumes that untrusted nodes can apply arbitrary measurements to their overall state, including collective measurements over multiple copies of the state. However, our setting does not correspond to a scenario involving an external eavesdropper or to a general information-theoretic security framework against arbitrary interception strategies, as considered elsewhere~\cite{Rosati2026,Kianvash2026}.

\subsection{Optimal states for precision and privacy}
In general, when seeking an optimal probe for a given estimation problem, it is necessary to recognize what the resource is. In the CV case, this can be subtle or nontrivial, as it may depend on: (i) The specific problem under consideration. For instance, one may be interested in noninvasive sensing, which requires the average or maximal received power at the nodes to be bounded. (ii) The class of states under consideration. For instance, multimode correlated Fock states differ substantially from Gaussian states, as the former are significantly more challenging to generate experimentally, and they are not robust to noise. In this case, the relevant resource may be identified through experimental constraints. (iii) The avoidance of pathological cases, in which a large Cramér-Rao bound does not correspond to a meaningful metrological protocol, as discussed below.

In DQS, the goal is to estimate the weighted average $\sum_{i=1}^iw_i\theta_i$. We have that
\begin{align}\label{CS}
\xi\leq\frac{w^T{\bf F}w}{(w^T\mathbb{P}_{\bf F}w)^2
}\overset{{\rm pure}}{=}\frac{4\text{Var}(G)}{\|w\|_2^4},
\end{align}
where ${G = \sum_{i=1}^M w_i G_i}$ is the generator~\cite{Proctor2018,Oh2020,oh2022distributed}, we have defined the variance ${{\rm Var}(G)=\langle G^2\rangle-\langle G\rangle^2}$, and ${\mathbb{P}_{\bf F}}$ is the projector on the range of ${\bf F}$. The first inequality in Eq.~\eqref{CS} is Cauchy-Schwarz applied to the vectors $\sqrt{{\bf F}}w$ and $\sqrt{{\bf F}^{-1}}w$,
and becomes an equality only if $\sqrt{{\bf F}} w \propto \sqrt{{\bf F}^{-1}} w$. This is indeed the case, for example, if ${\bf F}$ is rank-one, or if the state is fully symmetric and all weights are equal. The second equality in Eq.~\eqref{CS} holds for pure states. Moreover, we work with the ansatz that $w$ is in the range of $\mathbf{F}$, implying
${w^T \mathbb{P}_{\mathbf{F}} w = \|w\|_2^2}$.
This condition is required for the saturability of the Cramér-Rao bound
in Eq.~\eqref{CRbound}. 

When fixing the total average number of photons ${\sum_{i=1}^M \langle a_i^\dag a_i \rangle=N_{\rm tot}}$ in the probe, there exist known pathological cases where the achievable precision seems to diverge with finite resources. For instance, consider the state
\begin{align}\label{patholog}
|\psi\rangle = \sqrt{1-\delta}\,|0\rangle^{\otimes M} + \sqrt{\delta}\,\bigotimes_{k=1}^M|w_kN_{\rm tot}/\delta\rangle,
\end{align}
with ${0 < \delta < 1}$, chosen such that ${N_{\rm tot}/(M\delta)}$ is an integer.
The variance of the generator computed on $|\psi\rangle$ is 
\({\text{Var}(G)=N_{\rm tot}^2\|w\|_2^4\left(1/\delta-1\right)}\),
which can be arbitrarily large. The corresponding QFIM is given by ${{\bf F}=\left[4{\rm Var}(G)/\|w\|_2^4\right] ww^T}$, meaning that the privacy parameter is $1$. Notice that $w$ is in the range of ${\bf F}$, so the Cramér-Rao bound is in principle saturable. Since the QFIM has rank one, the inequality in Eq.~\eqref{CS} is saturated, and we have $\xi=4N_{\rm tot}^2(1/\delta-1)$. Since $\delta$ can be chosen arbitrarily small, it may seem that one could achieve infinite precision with finite resources, but this is not the case. In this example, the required prior variance is much smaller than $\xi$, so there is not a meaningful local metrological protocol achieving that precision~\footnote{This can be checked by considering the fidelity  ${F=|\langle \psi| e^{-i\phi G} |\psi\rangle|^2 = 1-2\delta (1-\delta) \left[1-\cos\left(\Omega\phi\right)\right]}$, where we have introduced ${\Omega=N_{\rm tot}\|w\|_2^2/\delta}$. One needs $\phi\lesssim \Omega^{-1}$ to stay within the same period, which translates to ${\phi^2\lesssim O(\delta^2/N_{\rm tot}^2)}$. This means that the prior variance is much smaller than $\xi$ when ${\delta\ll1}$.}.

Another possible constraint is to fix the maximum photon number, i.e., to restrict the state support to Fock states $\{|n\rangle\}$ with ${n \leq N_{\rm max}}$.  This effectively cuts the local Hilbert spaces. Under this restriction, the precision can no longer diverge, as the variance of $G$ is bounded.

Since we are interested in states that maximize privacy, we can restrict our attention to states for which ${\bf F} \propto w w^T$. Let us consider the state 
\begin{align}\label{optpsi}
|\psi\rangle=\sum_{n=0}^L \sqrt{p_n} \bigotimes_{k=1}^M |Kw_{k}n\rangle,
\end{align}
where we assume that all $w_{k}$ are rational numbers and $K$ is the least common multiple of the denominators of $w_k$ after reducing them to lowest terms, the sum over $n$ runs up to ${L=N_{\rm max}/(K \|w\|_\infty)}$, and the state is normalized as ${\sum_n p_n =1}$. Here, we get that ${{\rm Var}(G)=K^2\left[\sum_n p_n n^2-(\sum_{n}p_n n)^2\right]\|w\|_2^4/4}$, and the QFIM is ${{\bf F} = 4 {\rm Var}(G)ww^T/\|w\|_2^4}$, which ensures that the privacy parameter is $1$. One can maximize sensitivity by choosing ${p_0=p_L=1/2}$, and we get ${{\rm Var}(G) = N_{\rm max}^2} \|w\|_2^4/(2\|w\|_\infty)^2$. This brings to the following result.
\begin{theo}
Consider the problem of estimating ${\sum_{i}w_i\theta_i}$, where $\theta_i$ are local phase shifts generated by $G_i=a^\dag_i a_i$, and ${\sum_iw_i=1}$ with ${w_i>0}$. Given a constraint on the maximal photon number state, perfect privacy can be achieved by the states in Eq.~\eqref{optpsi} for any choice of $\{p_n\}$. The maximal precision is obtained by setting ${p_0 = p_L = 1/2}$, and is given by
\begin{align}
\xi = \left( \frac{N_{\rm max}}{\|w\|_\infty}\right)^{2}.
\end{align}
\end{theo}

For instance, if the weights are all equal, we have ${\xi= (MN_{\rm max})^2}$, which is optimal, as one can see using the inequality ${4{\rm Var}(G)\leq N_{\rm max}^2}$ in Eq.~\eqref{CS}. Therefore, perfect privacy can in principle also be achieved in the CV regime, with a precision that scales quadratically with the available resources. This scaling is optimal as the variance of $G$ can be at most $\mathcal{O}(N_{\rm max}^2)$.

The states required to reach this limit are \emph{multimode photon-number correlated states}, which are challenging to generate experimentally, as they rely on strong nonlinear interactions. Circuit-QED architectures, which naturally provide strong nonlinearities, have recently demonstrated the generation of such states for three modes, and feasible extensions to genuinely multimode settings have been proposed~\cite{Wilson2020}. However, in the presence of noise, the precision achievable with these states is rapidly reduced. In optical frequencies, these states can be prepared with postselection~\cite{schadow2026certification}, but this may lead to very low success probability for large networks. By comparison, Gaussian states are significantly easier to prepare in most experimental platforms, as they require at most two-photon interactions. Moreover, they are more robust to noise, making them compelling to analyze in the context of private DQS.

\section{Distributed quantum sensing with Fully-symmetric Gaussian states}
\subsection{Gaussian formalism}
Gaussian states \(\rho\) are quantum states generated by Hamiltonians that are at most quadratic in the canonical operators \cite{AlessioSerafini}. As such, they are fully characterized by their first-moment vector \(d\) and covariance matrix \(\bf V\). For an \(M-\)mode Gaussian state, we define the vector of quadrature operators \(R=(x_1,p_1,x_2,p_2,...,x_M,p_M)^T\), so that the canonical commutation relations take the compact form \([R_j,\,R_k]=i\Omega_{jk}\). Here, the \(2M\times2M \) matrix \(\Omega=\bigotimes_{j=1}^{M}i\sigma_y\), with \(\sigma_y\) being the second Pauli matrix, is the symplectic form. Given this structure, the first-moments vector is defined as \({d={\rm Tr}(\rho R)}\), while the entries of the covariance matrix are given by \({V_{jk}={\rm Tr}(\rho\{R_j-d_j,R_k-d_k\})}\), where \(\{\cdot,\cdot\}\) denotes the anticommutator. The uncertainty relation for a Gaussian state is ensured if \({\bf V}+i\Omega\geq0\) \cite{AlessioSerafini}.

 In the following, we impose a constraint on the average number of photons. Unlike the state in Eq.~\eqref{patholog}, Gaussian states do not exhibit pathological behavior under this constraint, since their photon-number variance is bounded.

\subsection{Isothermal fully-symmetric Gaussian states}
In this work, we consider a particular class of Gaussian states, namely, the class of FSG states with null first moments and with a bounded total number of photons. These states are invariant under any permutation of two modes, and their covariance matrix can be written in terms of \(2\times2\) blocks as \cite{Adesso2004,Serafini2005,AlessioSerafini}
\begin{equation}\label{covariance}
    {\bf V}=\begin{pmatrix}
        \epsilon&\gamma&...&\gamma\\
        \gamma&\epsilon&...&\gamma\\
        \vdots&\vdots&\ddots&\vdots\\
        \gamma&\gamma&...&\epsilon
    \end{pmatrix},
\end{equation}
where \(\epsilon=\text{diag}(\epsilon_1,\epsilon_2)\) and \(\gamma=\text{diag}(\gamma_1,\gamma_2)\). Their symplectic eigenvalues are 
\begin{align}
\nu^-&=\sqrt{(\epsilon_1-\gamma_1)(\epsilon_2-\gamma_2)}
, \label{puritycondition1}\\ \nu^+&=\sqrt{(\epsilon_1+(M-1)\gamma_1)(\epsilon_2+(M-1)\gamma_2)}\,\label{puritycondition2},
\end{align}
where $\nu^+$ is non-degenerate and $\nu^-$ is {${(M-1)}$-times} degenerate~\cite{Adesso2004,Serafini2005}, see Appendix~\ref{Appsym} for further details. The state is {\it isothermal} if all symplectic eigenvalues of $ {\bf V}$ are the same~\cite{PhysRevA.100.012323,Monras2013}, and, in particular, is pure if the symplectic eigenvalues are $1$. In the following, we set ${\nu^{-}=\nu^{+}=1+2 n_{\rm th}}$, where $n_{\rm th}$ represents the thermal noise at the beginning of the state preparation. Isothermal FSG states with a fixed average number of photons $N_{\rm tot}=M(\epsilon_1+\epsilon_2-2)/4$ are therefore defined by a single free parameter, since four parameters are subject to three constraints. This free parameter can be tuned to optimize either privacy or precision.

\begin{figure}[t!]
    \centering
\includegraphics[width=.48\textwidth]{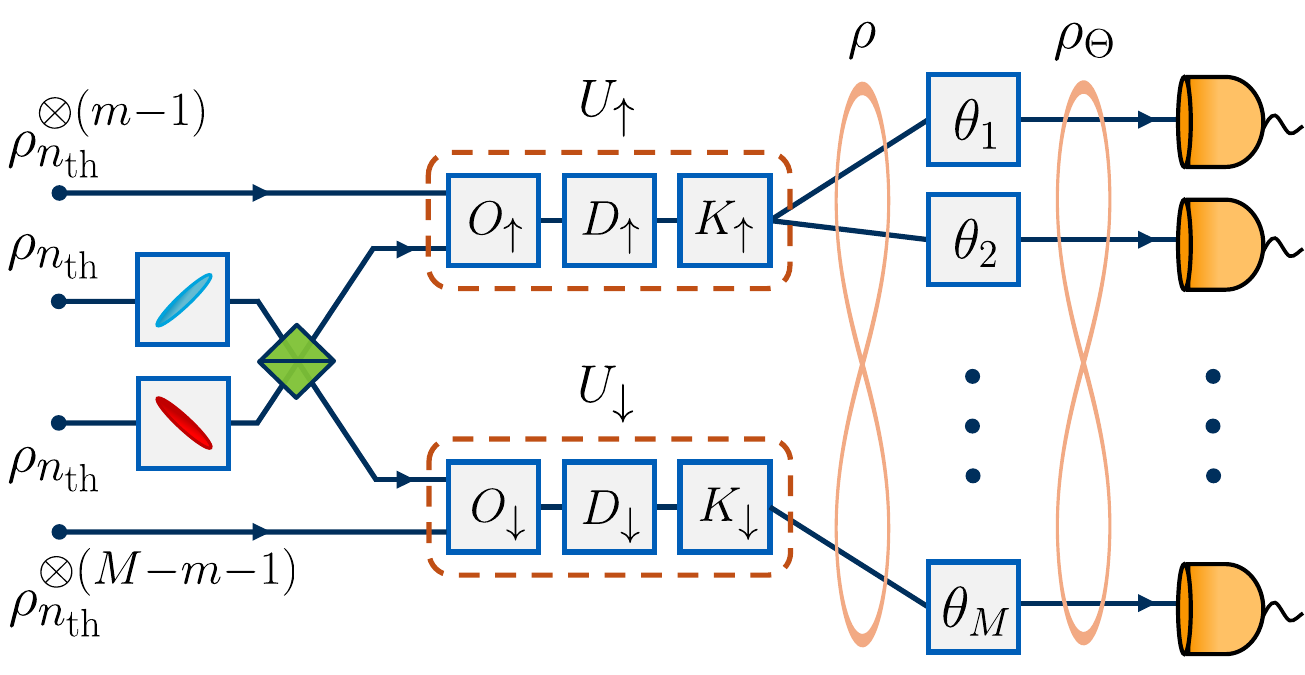}
    \caption{{\bf Scheme of a DQS protocol with a Gaussian quantum network.} Any $M$-mode fully-symmetric (permutation-invariant) Gaussian probe state $\rho$ can be prepared starting from $M$ thermal states $\rho_{n_{\rm th}}$. Two thermal modes are first converted into a two-mode squeezed state by means of orthogonal single-mode squeezing operations followed by a balanced beam splitter. In general, these single-mode squeezing operations may differ; they should be treated as equal when ${m = M/2}$. The upper (lower) output of the beamsplitter is then combined with $m-1$ (${M-m-1}$) thermal modes, where \(m\geq1\) can be chosen arbitrarily, and processed by a properly chosen Gaussian unitary $U_\uparrow$ ($U_{\downarrow}$)~\cite{Serafini2005}. By the Bloch--Messiah decomposition~\cite{AlessioSerafini,bloch1962canonical}, each Gaussian unitary can be decomposed as $U = ODK$, where $O$ and $K$ are passive transformations implementable using $\mathcal{O}(M)$ beam splitters, and $D$ implements a layer of single-mode squeezing on each mode. Each node then locally encodes a parameter \(\theta_i\) onto its share of the entangled state, in this paper implemented by a phase shift. 
Finally, each node performs a measurement and transmits the outcome to a central server, which processes the data to estimate a linear function of the local parameters.}
    \label{Fig2}
\end{figure}

Isothermal FSG states can be generated starting from a tensor product of \(M\) thermal states \(\rho_{n_{\rm th}}\) by means of passive linear optical transformations and single-mode squeezing operations~\cite{Serafini2005}, see Fig.~\ref{Fig2}. First, a two-mode squeezed state is generated by applying orthogonal single-mode squeezing to two thermal states, followed by a balanced beam-splitter operation. Then, one of the two output modes is combined with \(m-1\) thermal modes (with $m\geq1$) through a Gaussian unitary operation \(U_\uparrow\), while the other output mode is combined with the remaining \(M-m-1\) thermal modes through the Gaussian unitary operation \(U_\downarrow\). These unitary operations depend on the choice of $m$ (which is arbitrary), and they are defined constructively~\cite{Serafini2005}. By virtue of the Bloch–Messiah decomposition~\cite{bloch1962canonical,AlessioSerafini}, they can be implemented using two arrays of $\mathcal{O}(M)$ passive linear-optical transformations and one layer of single-mode squeezing operations on each mode.

 Notice that the covariance matrix of isothermal Gaussian states can be written as
${{\bf V}(n_{\rm th}) = (1+2n_{\rm th})\,{\bf V}(n_{\rm th}=0)}$.
This follows from the fact that the covariance matrix of a tensor product of thermal states at the same temperature $n_{\rm th}$ is
${{\bf V}_{\rm th} = (1+2n_{\rm th})\,\mathbb{I}_{2M}}$,
and that any symplectic transformation acting on this covariance matrix preserves the global multiplicative factor ${(1+2n_{\rm th})}$.  

\subsection{QFIM and privacy of isothermal FSG states}
For isothermal Gaussian states with null first moments, the entries of the QFIM depend on the encoded state covariance matrix \(\bf V_{\theta}\) as~\cite{Monras2013} 
\begin{equation}\label{MonrasQFIM}
    {\rm F}_{jk}=\frac{\nu^2}{2(1+\nu^2)}{\rm Tr}\left[{\bf V}_\theta^{-1}\left(\partial_{\theta_j}{\bf V}_\theta\right){\bf V}_\theta^{-1}\left(\partial_{\theta_k}{\bf V}_\theta\right)\right]\,,
\end{equation}
where \(\nu\equiv1+2n_{\rm th}\) is the symplectic eigenvalue.

For {\it pure} isothermal FSG states, Eq.~\eqref{MonrasQFIM} simplifies further and can be expressed in terms of the parameters of the probe state covariance matrix~\cite{Oh2020}: 
\begin{align}\label{QFIMentries}
{\rm F}_{jk}&=\begin{cases}
    \dfrac{\epsilon_1^2+\epsilon_2^2}{2}-1\equiv {\rm F}_{11}\,,&\text{for }j=k\\\\
    \dfrac{\gamma_1^2+\gamma_2^2}{2}\equiv {\rm F}_{12}\,, & \text{for }j\neq k.
\end{cases}
\end{align}
It follows that the QFIM takes the form
\({{\bf F}=({\rm F}_{11}-{\rm F}_{12})\mathbb{I}_{M}+{\rm F}_{12}\mathbb{J}_{M}}\)
where \(\mathbb{I}_{M}\) is the \(M\times M\) identity matrix, and \(\mathbb{J}_{M}\) an \(M\times M\) matrix with all entries equal to one.

Let us evaluate the Cramér–Rao bound in~\eqref{CRbound} for FSG states. The inverse of the QFIM has the form \({{\bf F}^{-1}=\alpha\mathbb{I}_M+\beta\mathbb{J}_M}\), with \({\alpha=({\rm F}_{11}-{\rm F}_{12})^{-1}}\) and \({\beta=[-\alpha{\rm F}_{12}]/[{\rm F}_{11}+(M-1){\rm F}_{12}]}\), as derived in Appendix \ref{AppInv}. Thus, we have
\begin{align}\label{xi}
    \xi^{-1}&=\sum_{j,k=1}^Mw_jw_k\left({\rm F}^{-1}\right)_{jk}\nonumber\\&=\|w\|_2^2\left[\left({\rm F}^{-1}\right)_{11}-\left({\rm F}^{-1}\right)_{12}\right]+\left({\rm F}^{-1}\right)_{12}\,,
\end{align}
where we used \(\|w\|_1^2=\sum_{j\neq k}^Mw_jw_k+\|w\|_2^2=1\). 

The privacy parameter in Eq.~\eqref{privacydefinition} can be straightforwardly estimated as 
\begin{equation}\label{privacydefinitionFSG}
    \mathcal{P}=\frac{\|w\|_2^2({\rm F}_{11}-{\rm F}_{12})+{\rm F}_{12}}{M\|w\|_2^2{\rm F}_{11}}\,.
\end{equation}

Given that all diagonal entries of the QFIM are equal, perfect privacy (i.e., ${\bf F}\propto ww^T$) is not achievable unless \(w_i=1/M\) for all \(i\in[1,M]\). This corresponds to the estimation of the mean function. This also implies that ${\rm F_{11}=F_{12}}$ is required for perfect privacy, since, for mean function estimation, $ww^T$ has all equal entries. 

In the following, we restrict the discussion to the case where all weights are equal, and find states that optimize estimation precision or privacy.

\begin{figure*}[t!]
    \centering
\includegraphics[width=.97\textwidth]{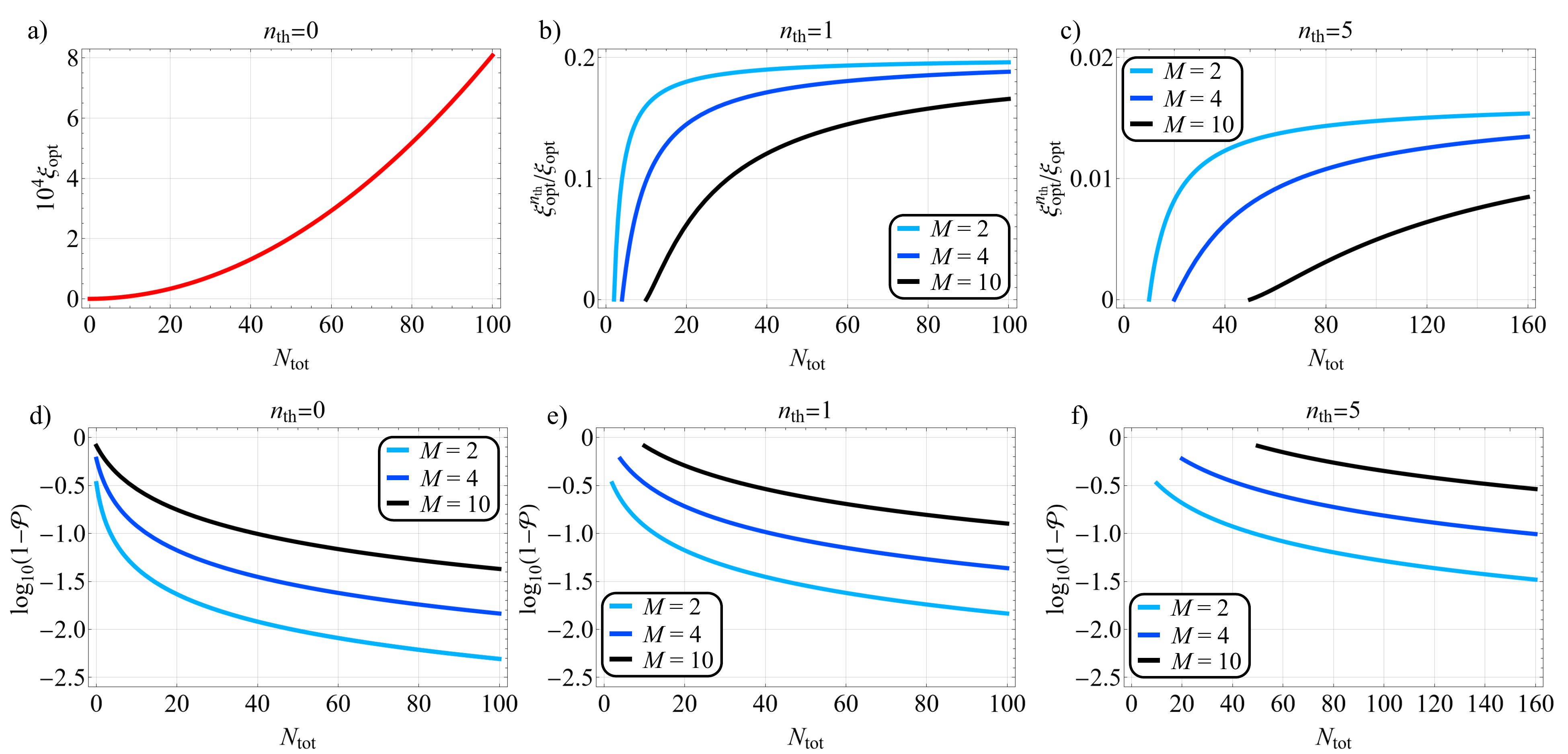}
    \caption{{\bf Isothermal FSG states maximizing estimation precision.} {\it Top}: \textbf{a)} optimal estimation precision ${\xi_{\rm opt}=8N_{\rm tot}(N_{\rm tot}+1)}$ for the estimation of the mean function as a function of the total number of photons $N_{\rm tot}$. This precision is achievable by the pure states given in Eq.~\eqref{epsilonottimo}. We also show the ratio between the optimal precision for mixed input states $\xi^{n_{\rm th}}_{\rm opt}$ and that for pure states $\xi_{\rm opt}$ for \textbf{b)} $n_{\rm th}=1$ and \textbf{c)} $n_{\rm th}=5$ as a function of \(N_{\rm tot}\), for different numbers of nodes $M$. For ${n_{\rm th}>0}$, $\xi_{\rm opt}^{n_{\rm th}}$ scales quadratically with $N_{\rm tot}$ as in the pure state case, since ${\xi_{\rm opt}^{n_{\rm th}}/\xi_{\rm opt}\sim 2/(1+\nu^2)}$ for sufficiently large $N_{\rm tot}$. 
{\it Bottom:} Privacy deficit $1-\mathcal{P}$ in semi-logarithmic scale for the states maximizing precision, for \textbf{d)} $n_{\rm th}=0$ (pure states), \textbf{e)} $n_{\rm th}=1$, and \textbf{f)} $n_{\rm th}=5$, and for different numbers of nodes $M$. For mixed states, privacy still approaches 1, but at a much slower rate.}
    \label{Fig3}
\end{figure*}

\subsection{Optimal precision of isothermal FSG states}
To obtain the FSG states with optimal precision for the estimation of the mean function, we need to optimize Eq.~\eqref{xi} with respect to the free parameter. The ultimate estimation precision considering Gaussian states as probe is $\xi_{\rm opt}=8N_{\rm tot}(N_{\rm tot}+1)$~\cite{Oh2020}. We check if a pure FSG state ($n_{\rm th}=0$) can achieve this precision by imposing
\begin{align}\label{optimalsensingcondition}
 \frac{\left({\rm F}^{-1}\right)_{11}-\left({\rm F}^{-1}\right)_{12}}{M}+\left({\rm F}^{-1}\right)_{12}=\frac{1}{8N_{\rm tot}(N_{\rm tot}+1)}.
\end{align}
Solving equation~\eqref{optimalsensingcondition} together with the purity and photon number constraints yields explicit expressions for the covariance matrix blocks \(\epsilon\) and \(\gamma\) depending on $N_{\rm tot}$ and $M$~\cite{Oh2020}:
\begin{align}
\gamma_{i}&=\frac{2N_{\rm tot}-2(-1)^i\sqrt{N_{\rm tot}(N_{\rm tot}+1)}}{M},\nonumber\\
    \epsilon_{i}&=1+\gamma_{i} \quad\quad\quad\quad \quad\quad\quad\,\,\,\,(i=1,2).\label{epsilonottimo}\,
\end{align}
The fact that it yields a physical state implies that the ultimate precision bound is achievable with pure FSG states. The privacy of such states is
\begin{equation}\label{privacyottimo}
   \mathcal{P}=1-\frac{M-1}{1+M+2N_{\rm tot}},
\end{equation}
which achieve $1$ only asymptotically for ${N_{\rm tot}\gg M}$. These results regarding the optimal precision and the corresponding privacy of pure FSG states are depicted in Fig.~\ref{Fig3}a and Fig.~\ref{Fig3}d, respectively. 

 Now, consider generic isothermal {\it Gaussian} states at temperature $n_{\rm th}$. These states have covariance matrix ${\bf V}(n_{\rm th})=(1+2n_{\rm th}){\bf V}(n_{\rm th}=0)$. The optimization within this set can be reduced to an optimization over the pure states set, by rescaling the total number of photons as ${N_{\rm tot} \rightarrow \frac{N_{\rm tot} - Mn_{\rm th} }{1+2n_{\rm th}}\equiv \tilde N_{\rm tot}}$, to satisfy the constraint on total number of photons.
In turn, the optimal estimation precision becomes ${\xi_{\rm opt} \rightarrow \frac{2\nu^2}{1+\nu^2}\xi_{\rm opt}(N_{\rm tot}\rightarrow\tilde N_{\rm tot})\equiv  \xi^{n_{\rm th}}_{\rm opt}}$, and it scales as
$ {\xi^{n_{\rm th}}_{\rm opt} \sim 16N_{\rm tot}^2/(1+\nu^2)}$
for ${N_{\rm tot} \gg M n_{\rm th}}$.
Thus, the quadratic scaling with the total number of photons is preserved, with a temperature-dependent multiplicative prefactor. Here and in the following, the superscript \(n_{\rm th}\) indicates estimation precision computed for \(n_{\rm th}>0\).
The privacy $\mathcal{P}$ can be obtained from Eq.~\eqref{privacyottimo} by exchanging ${N_{\rm tot}\rightarrow \tilde N_{\rm tot}}$, and it approaches unity only in the limit of infinite $N_{\rm tot}$.
The results for the states showing optimal precision and the corresponding privacy are given in Fig.~\ref{Fig3} for $n_{\rm th}=1$ and $5$. The deficit $1-\mathcal{P}$ increases by at most one order of magnitude when going from $n_{\rm th}=0$ to $n_{\rm th}=1$, whereas increasing $n_{\rm th}$ from $1$ to $5$ has a much smaller impact on privacy.

 Here, we have not considered the effect of displacement. In principle, one should maximize over displacement as well. However, in phase-shift estimation, the contribution to the precision arising from displacement scales only linearly with the number of photons associated with the displacement. This stands in contrast to the quadratic contribution of photons arising from correlation, as shown in Eq.~\eqref{optimalsensingcondition}. Consequently, for a sufficiently large number of photons, maintaining null displacement is the optimal choice.

\begin{figure*}[t!]
    \centering
\includegraphics[width=.97\textwidth]{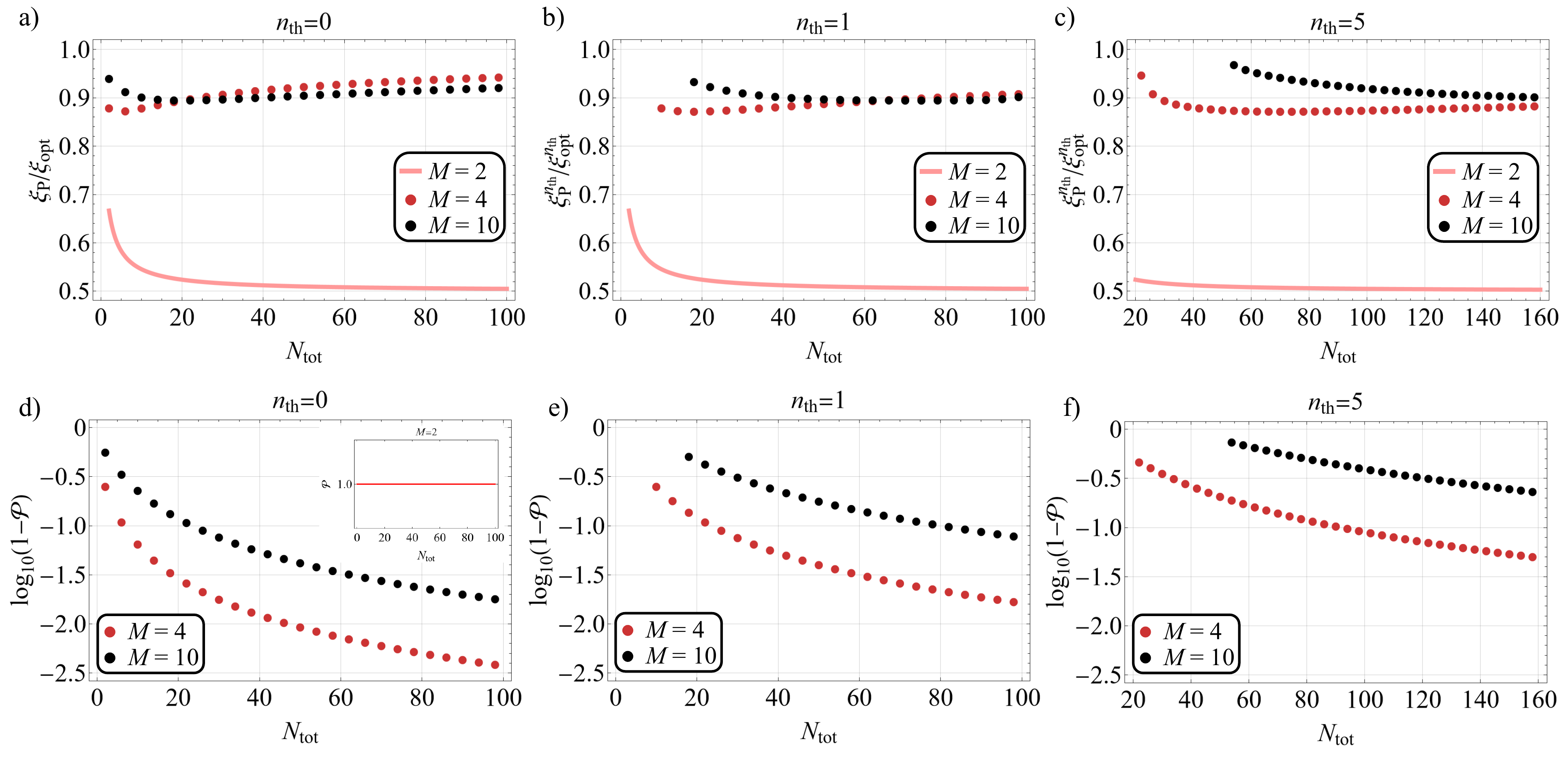}
   \caption{{\bf Isothermal FSG states maximizing privacy.} {\it Top}: Ratio between the estimation precision of FSG states maximizing privacy \(\xi_{\rm P}\) and that of FSG states maximizing precision \(\xi_{\rm top}\) (computed using Eq.~\eqref{xi}), as a function of the total number of photons $N_{\rm tot}$, for \textbf{a)} $n_{\rm th}=0$ (pure states), \textbf{b)} $n_{\rm th}=1$, and \textbf{c)} $n_{\rm th}=5$, and for different numbers of nodes $M$. When maximizing privacy, only a constant factor is lost in precision, which is close to $2$ for $M=2$ and close to $1$ for $M>2$. This implies that the quadratic scaling with $N_{\rm tot}$ is preserved if an optimal measurement is performed.
{\it Bottom:} Privacy deficit $1-\mathcal{P}$ in semi-logarithmic scale optimized over FSG states, for \textbf{d)} $n_{\rm th}=0$ (pure states), \textbf{e)} $n_{\rm th}=1$, and \textbf{f)} $n_{\rm th}=5$, and for different numbers of nodes $M$. While estimation precision is almost untouched, privacy deficit can improve considerably with respect to the results in Fig.~\ref{Fig3}. For instance, for $n_{\rm th}=0$, $M=4$, and $N_{\rm tot}=100$, we have $1-\mathcal{P}\simeq 10^{-2.42}$ in contrast with $10^{-1.83}$ that one gets in Eq.~\eqref{privacyottimo}. For \(M=2\), two-mode squeezed thermal states achieve perfect privacy for any \(N_{\rm tot}\) and \(n_{\rm th}\).}
    \label{Fig4}
\end{figure*}

\subsection{Optimal privacy of isothermal FSG states}
We now seek isothermal FSG states that maximize the privacy parameter. We have seen that for states maximizing precision, perfect privacy is achievable only in the limit of a large number of photons. In the following, we show that this feature is general for isothermal Gaussian states, except for the case ${M=2}$.

\begin{theo}
Consider the problem of estimating the mean function ${\frac{1}{M}\sum_{i}\theta_i}$, where $\theta_i$ are local phase shifts generated by $G_i=a^\dag_i a_i$. Among the $M$-modes isothermal FSG states at temperature $n_{\rm th}$, perfect privacy can be achieved only in two cases: (i) for ${M=2}$ by a two-mode squeezed thermal state, or (ii) in the limit ${N_{\rm tot} \rightarrow \infty}$ for a generic $M$ and $n_{\rm th}$.
\end{theo}
\begin{proof}
 Let us first consider the case \({n_{\rm th}=0}\). Since we seek ${{\bf F}\propto \mathbb{J}_M}$, and since for FSG states the QFIM takes the form \({{\bf F}=({\rm F}_{11}-{\rm F}_{12})\mathbb{I}_{M}+{\rm F}_{12}\mathbb{J}_{M}}\), perfect privacy requires \({\rm F_{11}=F_{12}}\), which is equivalent to ${C\equiv\epsilon_1^2-\gamma_1^2+\epsilon_2^2-\gamma_2^2-2=0}$. Taking the determinant of the covariance matrix in Eq.~\eqref{covariance}, after tracing out all but two modes, gives
${{(\epsilon_1^2 - \gamma_1^2)(\epsilon_2^2 - \gamma_2^2) \geq 1}}$,
as the symplectic eigenvalues of any physical Gaussian state are larger than or equal to $1$.  It follows that 
\begin{align}
C&\geq \epsilon_1^2-\gamma_1^2+\frac{1}{\epsilon_1^2-\gamma_1^2}-2 \nonumber\\
\quad &= \frac{(\epsilon_1^2-\gamma_1^2-1)^2}{\epsilon_1^2-\gamma_1^2}\geq0.
\end{align}
The first inequality is saturated only if the two modes are in a pure state. This implies that, for finite $N_{\rm tot}$, perfect privacy can be achieved only for ${M=2}$ using a pure state (${n_{\rm th}=0}$). A simple check shows that a TMSV state, for which 
${\epsilon_1 = \epsilon_2 = 1 + N_{\rm tot}}$ and 
${\gamma_1 = -\gamma_2 = \sqrt{N_{\rm tot}(N_{\rm tot}+2)}}$, 
achieves perfect privacy for any $N_{\rm tot}$. In fact, the TMSV state can be cast in the form of Eq.~\eqref{optpsi} in a certain limit.

We now turn to the case \({n_{\rm th}>0}\). For isothermal Gaussian states the covariance matrix obeys the relation \({{\bf V}(n_{\rm th})=(1+2n_{\rm th}){\bf V}(n_{\rm th}=0)}\). Using Eq.~\eqref{MonrasQFIM}, one can readily verify that the condition ${\rm F_{11}=F_{12}}$ holds if it is satisfied by ${{\bf V}(n_{\rm th}=0)}$.
The sole effect of \({n_{\rm th}>0}\) here is a rescaling of the total number of photons, \({N_{\rm tot}\to \frac{N_{\rm tot}-Mn_{\rm th}}{1+2n_{\rm th}}\equiv \tilde{N}_{\rm tot}}\) in ${{\bf V}(n_{\rm th}=0)}$, so that the total number of photons of ${{\bf V}(n_{\rm th})}$ is $N_{\rm tot}$. As a result, for finite \(N_{\rm tot}\) and \({n_{\rm th}>0}\), perfect privacy can be achieved only for \(M=2\), by a two-mode squeezed thermal state with parameters \({\epsilon_1=\epsilon_2=(1+N_{\rm tot})}\) and \({\gamma_1=-\gamma_2=\sqrt{N_{\rm tot}(N_{\rm tot}+2)-4n_{\rm th}(n_{\rm th}+1)}}\).

 If one allows for ${N_{\rm tot} \rightarrow \infty}$, perfect privacy becomes achievable for any $M$ and any temperature $n_{\rm th}$, as we has been shown in Eq.~\eqref{privacyottimo} for ${n_{\rm th}=0}$. This results holds for any $n_{\rm th}$, as explained in the discussion below Eq.~\eqref{privacyottimo}.
\end{proof}

Notice that for TMSV state, the achievable precision is \(\xi_{\rm TMSV}=4N_{\rm tot}(N_{\rm tot}+2)\). Remarkably, TMSV states achieve perfect privacy with only a factor of two loss in sensing precision compared to the optimal precision \(\xi_{\rm opt}\). We have not included displacement in our analysis. However, it is reasonable to anticipate that it offers no privacy advantage, since it conveys information about the local phases.

For generic \(M\) and \(n_{\rm th}\), we optimize the privacy expression in Eq.~\eqref{privacydefinitionFSG} with respect to the free parameter. As shown in Fig.~\ref{Fig4}, this optimization leads only to a small, constant reduction in precision: the multiplicative factor is \(1/2\) for \(M=2\), but is essentially unity for \(M>2\). At the same time, particularly in the pure-state case, this procedure significantly improves privacy, as quantified by the privacy deficit, compared to an unoptimized setting. In other words, states that optimize privacy retain near-optimal precision while driving privacy rapidly toward one for \(N_{\rm tot}\gg M\). For instance, for \(n_{\rm th}=0,\,M=4\), and \(N_{\rm tot}=100\), we obtain \(1-\mathcal{P}\simeq 10^{-2.42}\), in contrast with \(10^{-1.83}\) that one gets in Eq.~\eqref{privacyottimo}.

\begin{figure*}[t!]
    \centering
\includegraphics[width=.97\textwidth]{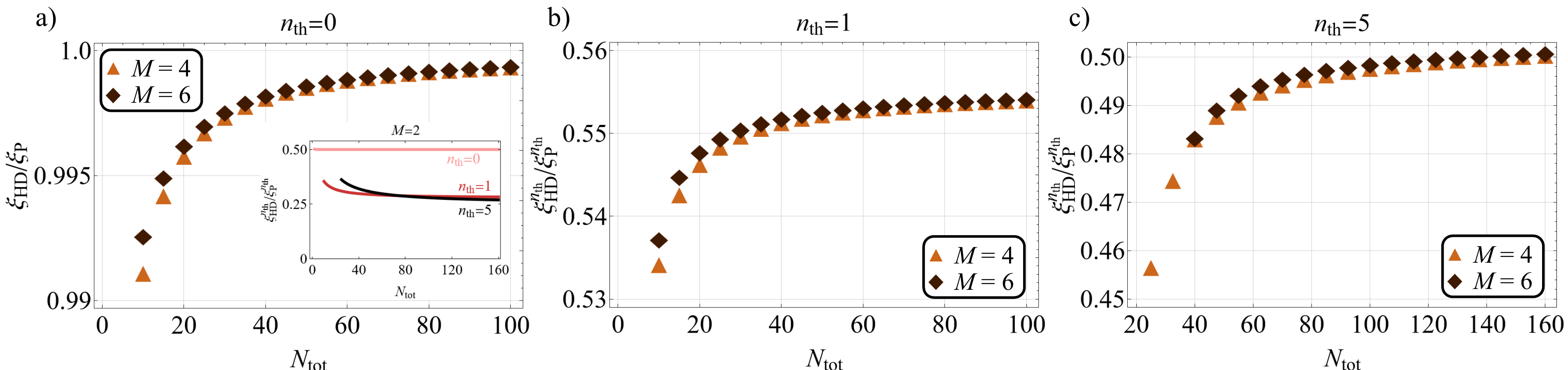}
    \caption{{\bf Precision achievable with homodyne detection.} Ratio between the estimation precision with homodyne detection \(\xi_{\rm HD}\) and the precision optimized over generic collective measurements \(\xi_{\rm P}\) (as shown in Fig.~\ref{Fig4}) on the privacy-optimized FSG states. Results are shown for \textbf{a)} $n_{\rm th}=0$ (pure states), \textbf{b)} $n_{\rm th}=1$, and \textbf{c)} $n_{\rm th}=5$, each for different numbers of nodes $M$. For \(M>2\) and pure states, homodyne detection is essentially optimal for sufficiently large \(N_{\rm tot}\), whereas for \(n_{\rm th}>0\) it becomes suboptimal, with the precision reduced by a factor \(\frac{1+\nu^2}{2\nu^2}\). The inset of panel {\bf a)} shows the same ratio for the two-mode squeezed state ($M=2$). Here, homodyne detection is already suboptimal for pure states (TMSV), with a factor 2 precision loss. For $n_{\rm th}>0$, the reduction becomes $\frac{1+\nu^2}{4\nu^2}$. }
    \label{Fig5}
\end{figure*}

\subsection{Local homodyne as optimal measurement}
So far, we have assessed precision performance through the QFIM. In general, saturating the corresponding Cramér–Rao bound requires a collective measurement across the nodes. However, a key requirement in DQS protocols is to employ local measurements on the nodes, particularly when aiming to ensure privacy. It is known that for pure FSG states maximizing precision—namely those in Eq.~\eqref{epsilonottimo}—a homodyne measurement with an optimized local phase already saturates the ultimate bound in Eq.~\eqref{CRbound}~\cite{Oh2020}. We now assess the performance of an optimized local homodyne measurement on the states that maximize privacy. The Fisher Information matrix for the homodyne detection is given by
\begin{align}
{\rm F}^{\rm (HD)}_{jk}=\frac{\text{Tr}\left\{{\bf \Gamma^{-1}}\left(\partial_{\theta_j}{\bf \Gamma}\right){\bf \Gamma^{-1}}\left(\partial_{\theta_k}{\bf \Gamma}\right)\right\}}{2}\,,
\end{align}
where \({\bf \Gamma}\) is the \({M\times M}\) homodyne covariance matrix with entries \({\Gamma}_{jj}=\epsilon_1\cos^2(\theta_{{\rm HD},j})+\epsilon_2\sin^2(\theta_{{\rm HD},j})\) and \({\Gamma_{jk}=\gamma_1\cos(\theta_{{\rm HD},j})\cos(\theta_{{\rm HD},k})+\gamma_2\sin(\theta_{{\rm HD},j})\sin(\theta_{{\rm HD},k})}\). Here, we have set the prior value of $\theta_j$ to zero. For symmetry reasons, we set $\theta_{\rm HD,i}=\theta_{\rm HD}$ for all $i$. Under these constraints, one can find the optimal angle \(\theta_{\rm opt,HD}\) by minimizing ${\rm Tr}(W{\bf F}^{\rm (HD)-1})$. Notice that the same angle also maximizes ${\rm Tr}(W{\bf F}^{\rm (HD)})$, which can be done numerically more easily (see Appendix~\ref{Apphomodyne} for details).

The estimation error for a locally 
unbiased estimator \(\hat{f}\) is bounded in this case by the classical Cramér-Rao bound~\cite{Paris2009}
\begin{equation}
    \Delta^2\hat{f}\geq\frac{w^T {\bf F}^{(\rm HD)-1}w}{n}\equiv \frac{1}{n\xi_{\rm HD}}\,,
\end{equation}
where \({\bf F}^{(\rm HD)}\) is the homodyne Fisher information matrix  and \(\xi_{\rm HD}\equiv (w^T{\bf F}^{(\rm HD)-1}w)^{-1}\) is the homodyne estimation precision. In Fig.~\ref{Fig5}, we plot the ratio between the estimation precision achievable with homodyne detection on the privacy-optimized FSG states \(\xi_{\rm HD}\) and the precision optimized over generic collective measurements \(\xi_{\rm P}\), as a function of the total number of photons \(N_{\rm tot}\).

 For \({M>2}\) and sufficiently large $N_{\rm tot}$, we find that in the pure state case (\({n_{\rm th}=0}\)), homodyne measurements essentially saturate the precision achievable by a general collective measurement, yielding \({\xi_{\rm HD}/\xi_{\rm P}\approx1}\). When \({n_{\rm th}>0}\), homodyne detection becomes sub-optimal; however, the corresponding estimation precision differs from that of a general collective measurement only by a constant factor \(\frac{1+\nu^2}{2\nu^2}\). This behavior can be traced back to the scaling of the homodyne covariance matrix for isothermal Gaussian states, \({{\bf \Gamma}(n_{\rm th})=(1+2n_{\rm th}){\bf \Gamma}(n_{\rm th}=0)}\). This implies that the precision for \(n_{\rm th}>0\) is obtained from the pure-state result by a simple rescaling of the total number of photons, i.e., \(\xi_{\rm HD}^{n_{\rm th}}=\xi_{\rm HD}(N_{\rm tot}\to\tilde{N}_{\rm tot})\). Moreover, since for isothermal Gaussian states \(\xi_{\rm P}^{n_{\rm th}}=\frac{2\nu^2}{1+\nu^2}\xi_{\rm P}(N_{\rm tot}\to\tilde{N}_{\rm tot})\) and since \(\xi_{\rm HD}/\xi_{\rm P}\approx1\) for enough large total number of photons, we conclude that \(\xi^{n_{\rm th}}_{\rm HD}/\xi_{\rm P}^{n_{\rm th}}\approx\frac{1+\nu^2}{2\nu^2}\). The case ${M=2}$ is exceptional: homodyne detection is suboptimal for pure states (TMSV), exhibiting a reduction in precision by a factor of \(2\). For \(n_{\rm th}>0\), we have instead \(\xi^{n_{\rm th}}_{\rm HD}/\xi_{\rm P}^{n_{\rm th}}\approx\frac{1+\nu^2}{4\nu^2}\). 

\begin{figure}[t!]
    \centering
\includegraphics[width=.48\textwidth]{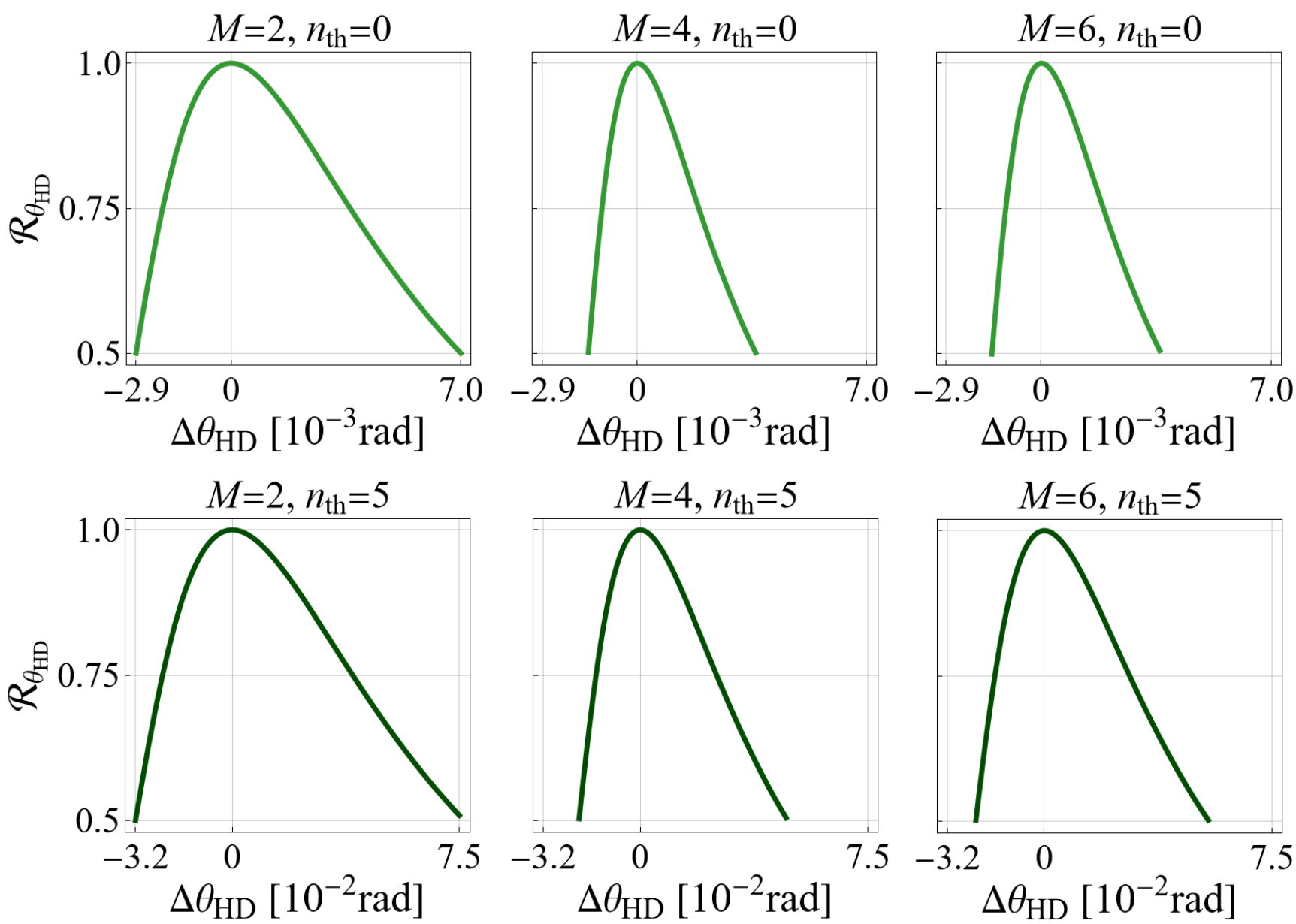}
    \caption{{\bf Homodyne angle sensitivity.} Ratio \(\mathcal{R}_{\theta_{\rm HD}}\) between the precision achieved by homodyne detection at an arbitrary angle $\theta_{\rm HD}$ and the precision attained at the optimal angle $\theta_{\rm opt,HD}$, shown as a function of the detuning \({\Delta\theta_{\rm HD}=\theta_{\rm opt,HD}-\theta_{\rm HD}}\). Pure probe states are more sensitive to deviations from the optimal angle than mixed states. The plots are obtained for \({N_{\rm tot}=100}\).
}
    \label{Fig6}
\end{figure}

In Fig.~\ref{Fig6}, we investigate the sensitivity of homodyne detection to deviations from the optimal measurement angle. We find that pure states are more sensitive to angle mismatches than mixed states. In particular, achieving optimal homodyne estimation precision for pure states requires calibrating the homodyne angle with a precision on the order of \(10^{-3}\)~rad, whereas for mixed states a precision of order \(10^{-2}\)~rad is sufficient. 

Achieving such angular precision requires a stable and well-defined phase reference between the probe field and the local oscillator employed for homodyne detection at each measurement node. In DQS, this translates into the necessity of a remote phase‑locking mechanism, similar to that used in continuous-variable quantum key distribution experiments~\cite{Cosmo2022}, which implies that a reference coherent signal must be transmitted from the server to each node. While technically challenging, such phase stability can be achieved over long distances in optical fibers~\cite{Droste2013, Zhang2020QKD} and over short distances in free space~\cite{Wolf2021}.

\section{Discussion}

 We have studied a DQS protocol in which each node encodes a parameter via a phase-shift operation. 
We have shown that multimode photon-number correlated states can simultaneously optimize precision and privacy, achieving perfect privacy in the ideal case. However, since their implementation is challenging and they are known to be highly sensitive to noise, we have focused on Gaussian probes, and in particular on isothermal FSG states. Our analysis shows that isothermal FSG states can achieve perfect privacy only asymptotically for a large number of photons $N_{\rm tot}$ when the number of nodes is ${M>2}$, or for any value of $N_{\rm tot}$ in the case of the two-mode squeezed thermal state (${M=2}$ nodes). Nevertheless, optimizing the states for privacy leads to only a minor reduction in estimation precision, while making the privacy converge rapidly to unity, as one can appreciate in Fig.~\ref{Fig4}.

Regarding the measurement, local homodyne detection is essentially optimal for pure states, whereas for mixed states it still preserves the quadratic scaling of precision with the total photon number. Concerning losses, we have modeled mixed states by adding thermal noise at the beginning of the preparation stage. The effect of losses—either during state preparation or after the beam splitter stage—remains to be analyzed, as they break the isothermal condition and require a dedicated study~\footnote{In Section IVB of Ref.~\cite{Oh2020}, the Authors use QFIM formulas valid for isothermal states~\cite{PhysRevA.100.012323}. However, they overlooked that adding amplitude-damping noise to an FSG state can result in a non-isothermal state.}. 

The analysis for Gaussian states has been focused on the equal weights case. However, the privacy level and the sensing precision of Gaussian quantum networks involving arbitrary linear combinations of parameters, and even generic functions of these parameters, can in principle be explored using recent results~\cite{Smerzi2023,Pezze2024,Jacobs2025,triggiani2021,triggiani2022}. Ultimately, although our analysis focuses on a specific class of Gaussian states, it is natural to expect that probes for optimally estimating the mean and achieving perfect privacy should be searched among the set of FSG states, given their symmetry properties. This would mean that perfect privacy with generic Gaussian states is not achievable with finite resources. However, this statement remains unproven in the most general Gaussian setting.

Our results pave the way for the development of practical, private distributed quantum sensing protocols in continuous-variable quantum networks. Given the passive nature of phase-shift operations, our work has potential applications in networks of low-energy devices, such as backscatter communication~\cite{jantti2017,DiCandia2018Backscatter,DiCandiaPRXQuantum,Zhang2024Backscatter} and ultra-low-power Internet-of-Things systems~\cite{wentzloff2020}.

{\it Relation with other works---} While finalizing this work, we became aware
of the related paper by A. de Oliveira Junior et al., arXiv:2509.12338~\cite{Oliveira2025privacy}, addressing the privacy question in continuous-variable DQS. Our analysis is complementary to theirs, as we focus on isothermal fully-symmetric Gaussian states, whereas they consider a specific class of bisymmetric Gaussian states~\cite{Serafini2005}. 

Their states consist of two blocks of modes, each of dimension \(M/2\) (with \(M\) required to be a power of \(2\) for their construction). They are invariant under any permutation of modes within the same block. Such states are generated as in Fig.~\ref{Fig2} with ${m=M/2}$ and ${D_{\uparrow\downarrow}=\mathbb{I}_{M}}$, i.e., no squeezing is applied at each side, but just a passive linear optical network of balanced beamsplitters. Their state becomes fully symmetric in the limit of a large number of photons.

Our approach requires additional experimental resources, as it introduces squeezing at the level of each block of modes. A more general framework could consider a generic bisymmetric state, encompassing both FSG states and the state of Ref.~\cite{Oliveira2025privacy} as limiting cases.

Compared to their probes, our optimized states exhibit a faster convergence of privacy toward unity for ${N_{\rm tot} \gg M}$. For example, for ${n_{\rm th}=0}$, ${M=4}$, and ${N_{\rm tot}=100}$, we get ${1-\mathcal{P}\simeq 4\times 10^{-3}}$, whereas the protocol in \cite{Oliveira2025privacy} yields ${1-\mathcal{P}\simeq 2\times 10^{-2}}$. For the estimation of the mean, compared to Ref.~\cite{Oliveira2025privacy}, we obtain a gain in precision by a factor ${2(1 + N_{\rm tot})/(2 + N_{\rm tot})}$ (which approaches $2$ in the limit ${N_{\rm tot} \gg 1}$), when considering the pure state that optimizes the precision~\footnote{In Ref.~\cite{Oliveira2025privacy}, the authors report only the maximal eigenvalue of the QFIM, $\lambda_{\rm max}$. The quantity $M \lambda_{\rm max}$ provides an upper bound on the precision for estimating the mean value of the parameters, as one can see from Eq.~\eqref{CS}, and we have compared our results with this upper bound.}. For the states optimized for privacy and with a local homodyne detection, the gain in precision is roughly unchanged, as shown in Figs.~\ref{Fig4}-\ref{Fig5}. Lastly, we show that for pure states homodyne detection is essentially optimal, except in the case $M=2$, where the precision is reduced to one half of the maximum. In contrast, Ref.~\cite{Oliveira2025privacy} does not establish the optimality of any local measurement for their setting, and explicitly identifies a nonlocal measurement saturating the Cramér-Rao bound only for $M=2$.

{\it Funding---} This study was funded by Academy of Finland, grants no. 349199, 353832, and 368477. 

\bibliography{Reply_Biblio.bib}

\appendix

\section{Determinant of the Block Covariance Matrix}\label{Appsym}
In this section, we show how to compute the determinant, and consequently the symplectic eigenvalues, of the covariance matrix in Eq.~\eqref{covariance}. This can be easily done by generalizing the results of Refs.~\cite{Adesso2004,Serafini2005}. The considered covariance matrix may be written using Kronecker products as
\begin{equation}
    {\bf V} = \mathbb{I}_{ M}\otimes(\epsilon-\gamma) + \mathbb{J}_{ M}\otimes\gamma\,,
\end{equation}
where \(\epsilon=\text{diag}(\epsilon_1,\epsilon_2)\), \(\gamma=\text{diag}(\gamma_1,\gamma_2)\), $\mathbb{I}_{ M}$ is the $M\times M$ identity and $\mathbb{J}_{ M}$ the $M\times M$ all-ones matrix. Notice that $\mathbb{J}_{ M}$ is a real symmetric matrix with eigenvalues $M$, (algebraic multiplicity $1$) and $0$ (algebraic multiplicity $M-1$). Thus, there exists an orthogonal transformation $U$ that diagonalizes $\mathbb{J}_{M}$. We can use \(U\) to build a transformation that brings ${\bf V}$ into a block-diagonal form. Indeed, if we define the orthogonal transformation $P=U\otimes \mathbb{I}_{2}$, we have
\begin{align}
P^T{\bf V}P&=P^T\left(\mathbb{I}_{ M}\otimes(\epsilon-\gamma) + \mathbb{J}_{ M}\otimes\gamma\right)P=\nonumber\\&=\mathbb{I}_{ M}\otimes(\epsilon-\gamma)+\text{diag}(M,\dots,0,0)\otimes\gamma\,.\label{blockdiag}
\end{align}
Using the determinant properties with respect to matrix multiplication, we have \(\det({\bf V})=\det(P{\bf V}_BP^T)=\det(P)\det({\bf V}_B)\det(P^T)=\det({\bf V}_B)\), where \({\bf V}_B\) is the block diagonal matrix \eqref{blockdiag}.
Substituting the explicit diagonal forms of $\epsilon$ and $\gamma$ in \eqref{blockdiag} yields the closed form
\begin{align}
    \det({\bf V})=&\left(\epsilon_{1}+(M-1)\gamma_{1}\right)\left(\epsilon_{2}+(M-1)\gamma_{2}\right)\times\nonumber\\&\left[(\epsilon_{1}-\gamma_{1})(\epsilon_{2}-\gamma_{2})\right]^{M-1}\,.
\end{align}
Ultimately, since for Gaussian states the determinant is simply the product of the symplectic eigenvalues squared~\cite{AlessioSerafini}, we retrieve the results in Eqs.~\eqref{puritycondition1}-\eqref{puritycondition2}.

\section{Inverse of the QFIM}\label{AppInv}
In this section, we compute the inverse of the QFIM and express it in terms of the entries of \(\bf F\). As mentioned in the main text, due to the symmetry of FSG states, the QFIM can be written as
\begin{equation}\label{QFIM structure}
{\bf F}=({\rm F}_{11}-{\rm F}_{12})\mathbb{I}_{M}+{\rm F}_{12}\mathbb{J}_{M}\,,
\end{equation}
where \(\mathbb{I}_{M}\) is the \(M\times M\) identity matrix, and \(\mathbb{J}_{M}\) an \(M\times M\) matrix with all entries equal to one. To compute the inverse matrix, we assume it has the same structure as \(\bf F\), so we take the ansatz \({\bf F}^{-1}=\alpha\mathbb{I}_M+\beta\mathbb{J}_M\), with \(\alpha,\beta\in\mathbb{R}\). Then, we have that
\begin{align}
{\bf F }{\bf F}^{-1}&=\Big[({\rm F}_{11}-{\rm F}_{12})\mathbb{I}_{M}+{\rm F}_{12}\mathbb{J}_{M}\Big]\Big[\alpha\mathbb{I}_{M}+\beta \mathbb{J}_{M}\Big]\nonumber\\
    &=\alpha({\rm F}_{11}-{\rm F}_{12})\mathbb{I}_M\nonumber\\&+\Big[\beta({\rm F}_{11}-{\rm F}_{12})+\alpha{\rm F}_{12}+M\beta{\rm F}_{12}\Big]\mathbb{J}_{M}\,.
\end{align}
Since \({\bf F} {\bf F}^{-1}=\mathbb{I}_{M}\), we must impose \(\alpha=({\rm F}_{11}-{\rm F}_{12})^{-1}\) and \(\beta=[-\alpha{\rm F}_{12}]/[{\rm F}_{11}+(M-1){\rm F}_{12}]\). Thus, the inverse of the QFIM is
\begin{align}\label{QFIMinverse}
    {\bf F}^{-1}&=\frac{1}{{\rm F}_{11}-{\rm F}_{12}}\left[\mathbb{I}_M-\frac{{\rm F}_{12}}{{\rm F}_{11}+(M-1){\rm F}_{12}}\mathbb{J}_{M}\right]\,.
\end{align}
For \(\rm F_{11}=F_{12}\), the inverse of the QFIM is not well defined, and one should perform the Moore-Penrose inverse instead. 

\section{Optimizing homodyne angle for fully-symmetric Gaussian states}\label{Apphomodyne}

Let us show that, due to the symmetry of the problem, maximizing \(w^T{\bf F}^{\rm (HD)}w\) is equivalent to minimizing \(w^T{\bf F}^{\rm (HD)-1}w\). By assuming that the prior information on all phase parameters is identical and that all homodyne angles are equal, the homodyne Fisher information matrix (FIM) takes the form \({{\bf F}^{(\rm HD)}={\rm \left(F_{11}^{(\rm HD)}-F_{12}^{(\rm HD)}\right)}\mathbb{I}_{ M}+{\rm F_{12}^{(\rm HD)}}\mathbb{J}_{ M}=a\mathbb{I}_{ M}+b\mathbb{J}_{ M}}\), where we defined \({a=\rm F_{11}^{(\rm HD)}-F_{12}^{(\rm HD)}}\) and \(b=\rm F_{12}^{(\rm HD)}\). Its inverse can be computed using Eq.~\eqref{QFIMinverse}. We then have
\begin{align}
   w^T{\bf F}^{(\rm HD)-1}w&=\frac{\|w\|_2^2}{a}-\frac{b}{a(a
   +Mb)}\nonumber\\
   &=\frac{1}{M(a+Mb)}\,,
\end{align}
where we have used the fact that \(w^T\mathbb{J}_{ M}w=1\) when \(\|w\|_1=1\), and, for mean estimation, \(\|w\|_2^2=1/M\). On the other hand, for the homodyne FIM we have
\begin{align}
    w^T{\bf F}^{(\rm HD)}w=wa\|w\|_2^2+b=\frac{a+Mb}{M}\,.
\end{align}
Therefore, maximizing \(w^T{\bf F}^{(\rm HD)}w\) is equivalent to minimizing \(w^T{\bf F}^{(\rm HD)-1}w\).

\end{document}